\documentclass[12pt]{article}
\usepackage{amsmath,amssymb}
\usepackage{graphicx,color}
\numberwithin{equation}{section}
\usepackage{cite}
\usepackage{bm}
\usepackage{dcolumn}  
\newcommand{\be}{\begin{equation}}
\newcommand{\bea}{\begin{eqnarray}}
\newcommand{\eea}{\end{eqnarray}}
\newcommand{\ba}{\begin{array}}
\newcommand{\ea}{\end{array}}
\newcommand{\ee}{\end{equation}}

\expandafter\ifx\csname mathbbm\endcsname\relax

\else

\fi \textheight 22cm \textwidth 15cm \topmargin 1mm
\begin{document}
\begin{titlepage}
\hfill
\vbox{
    \halign{#\hfil         \cr
           IPM/P-2006/002 \cr
                      } 
      }  
\vspace*{20mm}
\begin{center}
{\Large {\bf Non-Supersymmetric Attractors and Entropy Function}\\
}

\vspace*{15mm}
\vspace*{1mm}
{Mohsen Alishahiha\footnote{alishah@theory.ipm.ac.ir} and  Hajar Ebrahim\footnote{ebrahimi@theory.ipm.ac.ir}}
\vspace*{1cm}

{\it  Institute for Studies in Theoretical Physics
and Mathematics (IPM)\\
P.O. Box 19395-5531, Tehran, Iran \\ \vspace{3mm}}

\vspace*{1cm}
\end{center}

\begin{abstract}
We study the entropy of non-supersymmetric extremal black holes which exhibit attractor 
mechanism by making use of the entropy function. This method, being simple, can be used to 
calculate corrections to the entropy due to higher order corrections to the action.
In particular we apply this method for five dimensional non-supersymmetric extremal black hole which carries two magnetic charges and find the $R^2$ corrections to the entropy. 
Using the behavior of the action evaluated for the extremal black hole near the horizon, 
we also present a simple expression for C-function corrected by higher order corrections.

\end{abstract}
\end{titlepage}

\section{Introduction}

One of the interesting features of the black hole in ${\cal N}=2$ supersymmetric theories  is the 
attractor mechanism  \cite{{Ferrara:1995ih},{Strominger:1996kf},{Ferrara:1996dd}}. In this
mechanism the values of different scalar fields at the horizon are determined entirely by the
charges carried by the black hole regardless of their values at asymptotic infinity.  
Another important feature of the ${\cal N}=2$ supersymmetric theories is  
that the theory can be described by a function so called {\em prepotential}. 
Actually by making use of this rich structure 
of the supersymmetric black hole, one can study different 
aspects of these black holes such as the entropy. 

In fact the entropy of 1/2-BPS black holes in ${\cal N}=2$ supersymmetric string theories in 
four dimensions has been calculated which is in agreement  with microscopic  counting of the
states of the corresponding brane configurations representing these black holes
\cite{{Maldacena:1997de},{deWit:1996ix},{LopesCardoso:1998wt},{LopesCardoso:2000qm},{Cardoso:2000fp}}.  
 
Recently it has been shown that the Legendre transformation of the
entropy with respect to the electric charges of the black hole is related to the prepotential.
This observation has led to a
conjecture relating the entropy to the partition function of topological 
string theory
\cite{{Ooguri:2004zv},{Dabholkar:2004yr},{Verlinde:2004ck},{Ooguri:2005vr},
{Dijkgraaf:2005bp},{Pestun:2005ni}}. 
It could also give a practical way to compute the entropy of the black hole.

So far we have considered only the cases where the theory is supersymmetric. 
One may wonder if 
there is similar structure for non-supersymmetric black holes as well. Actually it has been shown that 
the attractor mechanism can also work for non-supersymmetric extremal black holes
\cite{{Ferrara:1997tw},{Gibbons:1996af},{Goldstein:2005hq},{Kallosh:2005bj},{Kallosh:2005ax},{Tripathy:2005qp},
{Giryavets:2005nf},{Goldstein:2005rr}}. Having had the attractor mechanism one would like to know whether 
this can be used to study different aspects of non-supersymmetric black holes such as entropy. 

More recently it has been shown \cite{Sen:2005wa} that the similar structure as what we have in the 
supersymmetric black hole appears in the non-supersymmetric case as well. This observation has
provided a simple method to compute the entropy of a spherically symmetric extremal black hole 
in a theory of gravity coupled to abelian gauge fields as well as neutral scalar fields with 
arbitrary higher order derivative interactions. In this method one first defines the {\em entropy function}
which 
is a function of the parameters labeling the near horizon background. By extremizing this function with
respect to the parameters one can find the values of these  parameters in terms of the black hole
charges. Moreover the entropy of the black hole is equal to this function at 
the extremum\footnote{We note that similar function has also been found in the supersymmetric case in \cite{{Behrndt:1996jn},{Mohaupt:2005jd}}. Although in the supersymmetric case is was not called entropy
function, it has precisely the same preoperties.}.

To be precise let us consider an extremal $D$-dimensional black hole whose near horizon 
geometry is $AdS_2\times S^{D-2}$ and carries electric and magnetic charges. We have also 
several scalar fields. One can define the entropy function by the 
Lagrangian density evaluated for this background integrated over $(D-2)$-sphere. The
Legendre transformation of this function with respect to the electric charges, evaluated at the extremum,
 is equal to the entropy
of the black hole divided by $2\pi$. In this sense the entropy function plays the role of the
prepotential in the supersymmetric case. This method has been used to compute corrections to the entropy
of the different black holes
due to higher order corrections to the effective action \cite{{Sen:2005iz},{Prester:2005qs}}.
It is the aim of this article to further study this method for non-supersymmetric black holes
which exhibit attractor mechanism.

The organization of the paper is as follows. In section 2 we shall develop the procedure of 
evaluating the entropy function for those extremal black holes which exhibit the attractor mechanism.
Using this method we will reproduce the results of \cite{Goldstein:2005hq}. In section 3 we will 
compute corrections to the entropy due to higher order corrections to the action using
entropy function. We will consider a particular form of the higher order corrections, namely 
the Lovelock type action. In section 4 we shall consider a specific model in which the equations
can exactly be solved. In five dimensions this model can be treated as a toy model for 
the five dimensional black hole considered in \cite{Strominger:1996sh}. We will also
compute $R^2$ corrections to the entropy and find the same structure as that in the supersymmetric case
which has recently been computed\footnote{The entropy of 
5-dimensional black hole in the presence of higher order corrections  was also studied 
in \cite{{Nojiri:2001ae},{Cvetic:2001bk},{Nojiri:2002qn}}} \cite{Guica:2005ig}. 
The last section is devoted to the discussions where we present
different aspects of the models we are considering. In particular we see how one can naturally defined 
a C-function for the models we are studying and how this function gets corrections due to higher
order corrections to the action.

\section{Entropy function for non-supersymmetric attractor}

In this section we shall study extremal black hole solution in $D$-dimensions whose 
metric has the following form 
\be
ds^2=G_{\mu\nu}dx^\mu dx^\nu=-a^2(r)dt^2+\frac{dr^2}{a^2(r)}+b^2(r)d\Omega_{D-2}^2,
\label{metric}
\ee 
where $a^{2}(r)$ has double zero at horizon. This black hole solution will also 
be supported by some scalar fields as well as electric and magnetic fields.

One might think about this as a solution  of an effective action which typically could have the 
following form
\be
I=\frac{1}{\kappa_D^2}\int d^Dx\sqrt{-G}\left(R-2(\partial\phi_i)^2-f^{(e)}_{ab}(\phi_i)F^{(e)}_aF^{(e)}_{b}-
f^{(m)}_{\alpha\beta}(\phi_i)F^{(m)}_\alpha F^{(m)}_\beta+\cdots\right),
\label{act}
\ee
where dots stand for higher order corrections. Of course we note that,
in general, adding higher order terms to the action could change the shape of the metric, nevertheless
we will assume that the isometry of the metric remains unchanged though the metric components might
get corrections. The only point which is crucial for our consideration is that $a^2(r)$ has double
zero at the horizon. In other words, we are interested in the extremal black hole solution.

We would like to study the entropy one can associate with this black hole.  
These kinds of black holes and their properties including entropy have recently been studied
in \cite{Goldstein:2005hq} in the context of non-supersymmetric attractors where the authors showed that
these solutions exhibit the attractor mechanism by which several moduli fields are drawn to 
fixed values at the
horizon of the black hole regardless of the values they take at asymptotic infinity. 
It has also been shown that the entropy of these black hole is given in terms of these 
values at horizon.

Here we shall use another approach to find the entropy of these extremal black holes.  
Following \cite{Sen:2005wa} we will first define entropy function which leads to a simple method
to evaluate the entropy of the black hole. Being simple, this method can also be used to find the corrections
to the entropy when we are taking into account higher order corrections to the action. 

To proceed we shall use the general formula for the entropy in the presence of higher derivative terms
which has been studied in \cite{{Wald:1993nt},{Visser:1993nu},{Jacobson:1993vj},{Iyer:1994ys},{Jacobson:1994qe}}.
In our case, taking into account that the covariant derivative of the tensor fields are zero, we get
\be
S_{BH}=8\pi\int d^{D-2}x\sqrt{G_{D-2}}\;\frac{\partial{\cal L}}{\partial R_{rtrt}}
g_{rr}g_{tt}\;.
\ee 
As it was shown in \cite{Sen:2005wa} one may find a simple expression for the entropy of the black hole
by using a particular rescaling of the coordinates. To be specific let us consider the following extremal black 
hole solution of the action (\ref{act}) \cite{Goldstein:2005hq}
\be
ds^2=-a^2(r)dt^2+\frac{dr^2}{a^2(r)}+b^2(r)d\Omega_{D-2}^2,\;\;\;\;\;\;F^\alpha_{D-2}=
{p^\alpha}\sqrt{\omega_{D-2}},
\label{sol1}
\ee 
where $\omega_{D-2}$ is the determinant of unit $(D-2)$-sphere.
We also have a non-zero scalar filed $\phi$. 
 
To find the entropy function and thereby the entropy one considers an ansatz for the extremal black hole solution
at near horizon as follows
\be
ds^2=v_1\left(-a^2(r)dt^2+\frac{dr^2}{a^2(r)}\right)+v_2b^2(r)d\Omega_{D-2}^2,\;\;\;\;\;\;F^\alpha_{D-2}=
{p^\alpha}\sqrt{\omega_{D-2}},
\label{solution}
\ee 
We note that in our notation one has 
$R_{trtr}=- \frac{a^2(r)''}{2v_1}g_{tt}g_{rr}$, which can be used to write the entropy formula as the 
following
\be
S_{BH}=-2\pi\;\frac{2}{a^2(r)''}\int d^{D-2}x\sqrt{-G}\;\frac{\partial{\cal L}}{\partial R_{\mu\nu\rho\sigma}}
R_{\mu\nu\rho\sigma},
\ee 
for $\mu,\nu,\rho,\sigma=r,t$, where $a^2(r)''=\frac{d^2a^2(r)}{dr^2}$.

Let us define ${\cal L}_{\lambda}$ to be the same as the original Lagrangian except that each factor 
of $R_{rtrt}$ in the expression of the ${\cal L}$ is multiplied by a factor of $\lambda$.
Thus one can rewrite the above equation as 
\be
 S_{BH}=-2\pi \frac{2}{a^2(r)''}\int d^{D-2}x\sqrt{-G}\;
 \frac{\partial{\cal L}_{\lambda}}{\partial \lambda}\bigg{|}_{\lambda=1}=-\frac{4\pi}{a^2(r)''} 
 \frac{\partial f_{\lambda}(\vec{p},v_i,\phi)}{\partial \lambda}\bigg{|}_{\lambda=1},
\label{en}
\ee
where the function $f$ is defined by
\be
f(\vec{p},v_i,\phi)=\int d^{D-2}x\sqrt{-G}\;{\cal L},
\label{fun}
\ee
and the right hand side is evaluated in the near horizon geometry  (\ref{solution}). Here
we have assumed that the solution has a horizon at $r=r_H$ where $a^2(r_H)=a^2(r_H)'=0$ and
the above expression has ultimately to evaluate at $r=r_H$. On the other hand the other parameters 
such as $v_i$ can be found using their equations of motion.
We note, however, that $f$ is a function of $r$ which in our procedure 
it is taken as a fixed parameter. In fact as we shall see the $r$-dependence will drop in the final 
expression  for the entropy. Actually since we are interested in the solutions which 
exhibit attractor mechanism, near the horizon one may assume that $a\approx 1-(r_H/r)^{D-3}$.
Therefore near the horizon we can define a new coordinate $\hat{r}=r-r_H$ such that 
$a\approx \frac{(D-3)\hat{r}}{r_H}$. In this coordinates system the solution near the 
horizon is $AdS_2\times S^{D-2}$.

Using the same argument as \cite{Sen:2005wa} one can see that the function ${f}_{\lambda}$ must be
of the from $v_1 \tilde{f}_{\lambda}(\vec{p},\phi,v_2,\lambda v_1^{-1})$ and thus 
we arrive at
\be
\lambda\frac{\partial f_{\lambda}(\vec{p},\vec{v},\phi)}{\partial \lambda}
+v_1\frac{\partial f_{\lambda}(\vec{p},\vec{v},\phi)}{\partial v_1}
-f_{\lambda}(\vec{p},\vec{v},\phi)=0.
\label{eq}
\ee
Therefore setting $\lambda=1$ in (\ref{eq})
and using the equations of motion one can find an expression for the entropy in terms of the function $f$ as follows
\be
S_{BH}=-\frac{4\pi}{a^2(r)''}\; f.
\label{ENT}
\ee
Here the left hand side is evaluated at the near horizon solution when the parameters $v_i$ and $\phi$ are fixed
using their equations of motion  which can be given by extremizing $f$ with respect to the 
corresponding parameters
\be
\frac{\partial f}{\partial \phi}=0,\;\;\;\;\;\;\;\;\;\;\;
\frac{\partial f}{\partial v_i}=0.
\ee
Here we assume that these equations have a solution.

Actually this is a special case of what considered in \cite{Sen:2005wa} where the entropy is found to be
the Legendre transformation of function $f$, defined as (\ref{fun}), with respect to the electric charges. But in
our case since the solution carries only magnetic charge the first term in the Legendre transformation is zero.
We will come back to this point later. 

To see how this formalism works let us compute $f$ for the solution (\ref{solution}). Doing so 
we arrive at
\be
f=\frac{\Omega_{D-2}}{16\pi } v_1v_2^{\frac{(D-2)}{2}}b^{D-2}\left(\frac{(D-3)(D-2)v_1-2(D-3)^2v_2}{v_1v_2b^2}
-\frac{(D-2)! V_{\rm eff}}{v_2^{D-2}b^{2(D-2)}}\right),
\ee
where $V_{\rm eff}=f_{\alpha\beta}(\phi)p^\alpha p^\beta$.
In writing this equation we have used the fact that $a^2$ has double zero root at 
horizon and also $b^2(r)a^2(r)''=2(D-3)^2$ in the near horizon limit. 
The later assumption can be obtained directly from the equations of motion.
Now we need to extremize $f$ with respect to $v_i$ and $\phi$
\be
\frac{\partial f}{\partial v_1}=0,\;\;\;\;\;\; \frac{\partial f}{\partial v_2}=0,
\;\;\;\;\;\;\frac{\partial f}{\partial \phi}=0. 
\ee
The last equation leads to $\partial_\phi V_{\rm eff}=0$ which we assume that this equation as a solution
at $\phi=\phi_0$. From the first
two equations one finds
\be
v_1=v_2=\left(\frac{(D-4)!\; V_{\rm eff}}{b^{2(D-3)}}\right)^{\frac{1}{D-3}}.
\ee  
Evaluating $f$ at these values we get
\be
f=-2\frac{\Omega_{D-2}}{16\pi}\;\frac{(D-3)^2}{b^2}\bigg{(}(D-4)!\;V_{\rm eff}(\phi_0)\bigg{)}^{\frac{D-2}{2(D-3)}}.
\ee
Therefore the entropy is given by 
\be
S=\frac{\Omega_{D-2}}{4}\bigg{(}(D-4)!\;V_{\rm eff}(\phi_0)\bigg{)}^{\frac{D-2}{2(D-3)}}.
\ee
It is worth noting that the $r$-dependence is dropped in the final result and the entropy is 
given by the value of the potential at its extremum which is given in terms of the magnetic
charge and $f_{\alpha\beta}(\phi_0)$. We note also that by plugging $v_1$ and $v_2$ into the metric, one can read
the radius of horizon in terms of the effective potential  
\be
r_H^{2}=\bigg{(}(D-4)!\;V_{\rm eff}(\phi_0)\bigg{)}^{1/(D-3)},
\ee
in agreement with \cite{Goldstein:2005hq}.

\section{Higher order corrections to entropy function }

In this section we shall study corrections to the entropy due to higher order corrections to the action. 
In general adding higher order corrections can change the entropy in two
different ways. The corrections could be due to the additional terms in the action when we are 
evaluating $f$ using the zeroth order solution for the metric and other fields. One may also
consider the case where the modification is due to the fact that adding these terms would change
the equations of motion and thereby change the solution. 
We note that in the procedure we use to compute the entropy both effects will be 
taken into account.

To be specific we consider higher order corrections to the action to be of the Lovelock type
where the higher order corrections are given by the extended Gauss-Bonnet action \cite{Lovelock:1971yv}
\be
I=\frac{1}{\kappa^2}\int d^Dx\sqrt{G}\sum_{m=1}\frac{\lambda_m}{2^m}\;\delta^{\rho_1\sigma_1\cdots \rho_m\sigma_m}_{\mu_1\nu_1\cdots
\mu_m\nu_m}\;R^{\mu_1\nu_1}{}_{\rho_1\sigma_1}\cdots R^{\mu_m\nu_m}{}_{\rho_m\sigma_m}, 
\ee
where $\lambda_m$ are free parameters which could be either constant or one may consider the case where
they depend on the scalar fields. $\delta^{\rho_1\sigma_1\cdots \rho_m\sigma_m}_{\mu_1\nu_1\cdots
\mu_m\nu_m}$ is totally antisymmetric product of $m$ Kronecher deltas, normalized to take values $\pm 1$.
For $m=1$ setting $\lambda_1=1$ we get standard Einstein action, for $D=2m$ the $m$th term is topological
and for $D$ dimensional space-time all terms for $m>D/2$ identically are equal to zero.

The corrections to the entropy of the small black hole in the heterotic string theory due to this
action has recently been studied in \cite{Prester:2005qs} where the near horizon solution is 
$AdS_2\times S^{D-2}$. It is the aim of this section to study these corrections for the 
extremal black hole solution given by (\ref{sol1}) using 
the procedure we have developed in the previous section.  
  
To find the entropy we need to calculate function $f$ for the solution (\ref{solution}) using the
above action. Actually it was shown \cite{Cvitan:2002cs} that for the metric of type
\be
ds^2=g_{ab}dx^adx^b+r^2(x)d\Omega_{D-2}^2,\;\;\;\;\;\;\;a,b=1,2,
\label{gen}
\ee 
one gets the following expression for the Gauss-Bonnet densities integrated over the
unit $(D-2)$-sphere
\bea
\frac{1}{\kappa^2}\int d^{D-2}x\sqrt{-G}{\cal L}_m&=&-\frac{\Omega_{D-2}}{\kappa^2}
\frac{(D-2)!}{(D-2m)!}\lambda_m\sqrt{-g}\;r^{D-2m-2}[1-(\nabla r)^2]^{m-2}\cr 
&\times&\bigg{\{}2m(m-1)r^2[(\nabla_a\nabla_b r)^2-(\nabla^2 r)^2]\cr
&&\;\;+2m(B-2m)r\nabla^2r[1-(\nabla r)^2]-mr^2{\cal R}[1-(\nabla r)^2]\cr
&&\;\;-(D-2m)(D-2m-1)(1-(\nabla r)^2]^2\bigg{\}},
\label{gb}
\eea
where ${\cal R}$ is the 2-dimensional Ricci scalar of the metric $g_{ab}$. 
Plugging the solution (\ref{solution}) into the equation (\ref{gb}) and taking into account
that all terms which have covariant derivative vanish for the near horizon solution,
one can find $f$ as follows
\bea
f&=&\frac{\Omega_{D-2}v_1}{16\pi}\bigg{\{}-\frac{(D-2)!\;V_{\rm eff}}{v_2^{(D-2)/2}b^{D-2}}
+
\sum_{m=1}^{[D/2]}\frac{(D-2)!}{(D-2m)!}\lambda_m v_2^{(D-2m)/2}b^{D-2m}\cr
&&\;\;\;\;\;\;\;\;\;\;\;\;\;\;\;\;\;\;\times
\left(\frac{(D-2m)(D-2m-1)v_1-mv_2b^2a^2(r)''}{v_1v_2b^2}\right)\bigg{\}}.
\label{gb2}
\eea
For $m=1$ setting $\lambda_1=1$ we get the leading order action studied in the previous section, the other terms 
can be considered as higher order corrections to the action. We still assume that $a^2(r)$ has double zero 
root at the horizon which means that although it might get corrections from higher order terms in the action the
$a^2(r)''$ remains a non-zero constant at the horizon. 
Actually this is the only assumption we make in the rest of this section.
Now the same as before we will consider $b$ (or $r$) as a fixed parameter and extremize $f$ with 
respect to $v_i$ and $\phi$. Then the  entropy will be given by (\ref{ENT}) at this extremum.

To get an insight how this procedure works, let us study some examples explicitly. 
In four dimensions we have only $m=1,2$ and so we get
\be 
f=\frac{1}{4}\left(2v_1-v_2b^2a^2(r)''-\frac{2v_1}{v_2b^2}V_{\rm eff}\right)-\lambda_2 a^2(r)'',
\ee
from which one finds $v_1=a^2(r_H)''V_{\rm eff}(\phi_0)/2,\;v_2=V_{\rm eff}(\phi_0)/b^2$ and therefore we arrive at
\be
S=\pi V_{\rm eff}(\phi_0)+4\pi\lambda_2.
\ee 
Note that the entropy is corrected by a constant due to Gauss-Bonnet term.
 On the other hand using 
this result and plugging them into the metric one can read the radius of the horizon which is 
given by $r_H^2=V_{\rm eff}$ that is the same as the leading order which we have studied in the 
previous section. We note, however, that $\lambda_2$ could be a function of $\phi$, {\it i.e.}
$\lambda_2=g(\phi)\tilde{\lambda}_2$. In this case although the expression for $r_H$ in terms of the
effective potential is the same as the leading order, it could  get corrections due to the corrections
of $\phi_0$ which in this case it is the solution of 
$\partial_\phi V_{\rm eff}(\phi)=-4\tilde{\lambda}_2\partial_\phi g(\phi)$. 
 
In five dimensions we still have $m=1,2$ and the function $f$ reads
\be
f=\frac{\pi}{8}\left(b\sqrt{v_2}(6v_1-v_2b^2a^2(r)'')-\frac{6v_1}{v_2^{3/2}b^3}V_{\rm eff}\right)-\frac{3}{2}\pi\tilde{\lambda}_2g(\phi)\sqrt{v_2}ba^2(r)'',
\ee
which upon extremizing it with respect to $v_1$ and $v_2$ one gets 
\be
v_2=\frac{\sqrt{V_{\rm eff}(\phi_0)}}{b^2},\;\;\;\;\;\;\;\;\;\;v_1=\frac{a^2(r)''}{8}\left(\sqrt{V_{\rm eff}(\phi_0)}+4\tilde{\lambda}_2g(\phi_0)\right),
\ee
where $\phi_0$ is a solution of $\partial_\phi f=0$. The entropy is given by
\be
S=\frac{\pi^2}{2}\left(V_{\rm eff}^{3/4}(\phi_0)+12\tilde{\lambda}_2 g(\phi_0)V^{1/4}_{\rm eff}(\phi_0)\right).
\label{5bh}
\ee

It is worth noting that to get consistent results one needs to assume that $\partial_\phi f=0$ has a 
solution. Since at leading order the function $f$ is proportional to the effective potential it means that
the effective potential has to have an extremum. Taking into account the higher order corrections, with
$\phi$-dependent coefficients, the condition cannot be given only in terms of the effective potential and
in fact one will have to put the condition directly on $f$ which is the generalization of the
leading order condition. Namely the function $f$ has to have an extremum with respect to $\phi$, {\em i.e.}
$\partial_\phi f|_{\phi_0}=0$ must have a solution. Therefore as far as the entropy computation is
concerned we just need to extremize the entropy function and the entropy is given in terms of the value of entropy function at extremum.

We note, however, that the way the entropy function is defined might give an insight whether one needs to put another
condition on $f$. In fact the function $f$ is very similar to what is defined as the free energy for a system
with gravitational interaction \cite{Hawking:1982dh} in which one can identify the free energy with the
Euclidean gravitational action times the temperature. If $f$ could be interpreted as the free energy of 
the system, then one may want to assume $\partial_\phi^2 f|_{\phi_0}>0$ 
to get the stable solution where the free energy 
is minimum. Otherwise the solution could be unstable which might mean
there is not attractor behavior\footnote{Non-supersymmetric attractor
mechanism in the presence of $R^2$ correction has recently been studied in \cite{TT}.}.
Note that in leading order this condition is equivalent to 
have minimum for the effective potential which is crucial to have 
attractor mechanism \cite{Goldstein:2005hq}. 

\section{Exact solution}

Let us now consider an explicit example in which the equations can exactly be solved. To be
precise consider a system with one scalar and two gauge fields and the following potential
\be
V_{\rm eff}(\phi)=e^{\alpha_1\phi}p_1^2+e^{\alpha_2\phi}p_2^2.
\ee
For the 4-dimensional extremal black hole this potential have been studied in 
\cite{Goldstein:2005hq} where the authors have noticed that for the special case 
where $\alpha_1=-\alpha_2$ the entropy is given by
\be
S=2\pi|p_1p_2|.
\label{a2}
\ee
For $\alpha_1=-\alpha_2=2$ it is in fact the supersymmetric black hole solution 
studied in \cite{Kallosh:1992ii} whose entropy is exactly (\ref{a2}). Note that although for 
generic values of $\alpha_1$ and $\alpha_2$
the solution is not supersymmetric, as long as $\alpha_1=-\alpha_2$ 
the entropy is still given by (\ref{a2}). Therefore one might conclude that as far as
the entropy of the black hole is concerned it is the attractor
mechanism which plays the role which is the same for supersymmetric and non-supersymmetric cases.

Let us now consider higher order corrections to this entropy using the results we have
presented in the previous section. To be concrete we shall consider the case where
$g(\phi)=e^{\alpha_3\phi}$. In this case    
the critical value $\phi_0$ is given by the solution of the following equation
\be
e^{-2\alpha_2\phi_0}+\frac{4\tilde{\lambda}_2\alpha_3}{\alpha_2p_1^2}e^{(\alpha_3-\alpha_2)\phi_0}
-\frac{p_2^2}{p_1^2}=0\;.
\ee
To proceed let us further assume that $\alpha_3=\alpha_2$. For this case the entropy is given by
\be
S=\pi |p_1p_2|\left((1-\frac{4\tilde{\lambda}}{p_2^2})^{1/2}+ 
(1-\frac{4\tilde{\lambda}}{p_2^2})^{-1/2}\right)+4\pi\tilde{\lambda}\bigg{|}\frac{p_1}{p_2}\bigg{|}
(1-\frac{4\tilde{\lambda}}{p_2^2})^{-1/2}.
\ee
In the large $p_2,p_1$ limit, one may expand this expression to get
\be
S=2\pi |p_1p_2|+4\pi\tilde{\lambda}\bigg{|}\frac{p_1}{p_2}\bigg{|}+{\cal O}( p_2^{-2}).
\ee
We note that if we had considered $\alpha_3=-\alpha_2$, we would have gotten the same result as above
expect that the leading order is now proportional to $|p_2/p_1|$.

Let us also study the five dimensional extremal black hole with the same potential as above, though 
here we will consider the case where $\alpha_1=-2\alpha_2$. We shall first evaluate the entropy in the 
leading order which is given in terms of the effective potential $V_{\rm eff}(\phi_0)$ where $\phi_0$ is
given by
\be
e^{\alpha_2\phi_0}=\left(\frac{\sqrt{2}p_1}{p_2}\right)^{2/3},
\ee
and therefore we arrive at 
\be
S=\frac{3^{3/4}\pi^2}{2}\sqrt{\frac{p_1p_2^2}{2}}.
\ee
This could be compared with 5-dimensional black hole studied in \cite{Strominger:1996sh},
of course we have an extra $3^{3/4}\pi/4$ factor. Indeed as far as the potential is concerned this case has the same potential structure as that considered
in \cite{Strominger:1996sh}. Therefore this model could be used to understand the five dimensional
black hole better. We will come back to this point later.

On the other hand adding the higher order correction to the
action such that $\lambda_2=\tilde{\lambda}_2e^{\alpha_3\phi}$ and using the correction we found 
for the entropy for the 5-dimensional black hole (\ref{5bh}) and setting $\alpha_3=\alpha_2$ one finds
\be
S=(\frac{3^{3/4}\pi}{4})\left(2{\pi}\sqrt{\frac{p_1p_2^2}{2}}\right)\left[1+2^{8/3}\sqrt{3}\tilde{\lambda}
\left(\frac{p_1}{p_2^{4}}\right)^{1/3}+\cdots
\right].
\ee

Higher order corrections for 5-dimensional black holes and black rings have recently been studied in 
\cite{Guica:2005ig}. Of course in this paper the authors have considered BPS solution, though in our 
case we have not assumed any supersymmetry. It is worth noting that although our solution is not
supersymmetric, we get a correction to the entropy which has the same structure as that in BPS black
hole studied in \cite{Guica:2005ig}, namely
setting $p_1=2p_2=2Q$, the above equation reads
\be
S=(\frac{3^{3/4}\pi}{4})(2{\pi}\sqrt{Q^3})\left(1+2^{3}\sqrt{3}
\frac{\tilde{\lambda}}{Q}+\cdots\right).
\ee
It would be interesting to further study this case and compare this non-supersymmetric model with 
the BPS black hole solution.
 
In general we can consider $D$-dimensional black hole solution with the above effective
potential. In this case assuming $\alpha_1=-(D-3)\alpha_2$, in leading order, one finds
\be
S\sim (p_1p_2^{D-3})^{1/(D-3)}.
\ee
 
\section{Discussions}

In this section we shall study different physical aspects of the black hole solutions
we have considered in the previous sections. So far we have studied higher order 
corrections to the entropy of extremal black hole
(\ref{solution}) by making use of the entropy function which can be defined as the Lagrangian 
density integrated over $(D-2)$-sphere. In this procedure one only needs to know the behavior of 
the solution near the horizon. Although the Lagrangian density is in general $r$-dependent function,
the $r$ is treated as a fixed parameter. Of course it will dropped in the final expression we have 
found for the entropy. In fact since the physics is governed by the
behavior of the black hole solution in near horizon limit, one should set $r=r_H$  which can be
found in terms of the effective potential in its extremum.

Of course it is not the only way to calculate the higher order corrections to the entropy of an extremal 
black hole with spherically symmetric metric. Actually it has been shown \cite{Solodukhin:1998tc}
(see also \cite{Cvitan:2002cs}) that for general relativity 
in $D\geq 3$ dimensions one can find an effective two dimensional theory which governs the conformal dynamics at the
horizon and the entropy of the black hole is given by the central charge of the corresponding Virasoro algebra.
It is interesting to compare this computations with what we have done in this paper.

To do that we start from the most general action we have considered in this paper 
and evaluate it for the ansatz (\ref{gen}) which leads to an effective two dimensional action
such that at near horizon one finds 
\bea
I&=&\frac{\Omega_{D-2}}{16\pi }\int d^2x\sqrt{-g}\bigg{\{}-\frac{(D-2)!\;V_{\rm eff}}{b^{D-2}}
+\sum_{m=1}^{[D/2]}\frac{(D-2)!}{(D-2m)!}\lambda_m b^{D-2m}\cr
&&\;\;\;\;\;\;\;\;\;\;\;\;\;\;\;\;\;\;\;\;\;\;\;\;\;\;\;\;\;\;\;\times
\left(\frac{(D-2m)(D-2m-1)-mb^2{\cal R}}{b^2}\right)\bigg{\}}.
\eea
Here we have used the fact that the covariant derivative of different fields goes to zero near horizon.
Following  \cite{{Solodukhin:1998tc},{Cvitan:2002cs}} if we define 
\be
\varphi=\frac{2\Phi^2}{q\Phi_H},\;\;\;\;\;\;\tilde{g}_{ab}=\frac{d\varphi}{d b} e^{-2\varphi/q\Phi_H}g_{ab},
\;\;\;\;\;\Phi^2=2\frac{\Omega_{D-2}}{\kappa^2_D}\sum_{m=1}^{[D/2]}m\lambda_m\frac{(D-2)!}{(D-2m)!}\;b^{D-2m},
\ee
the above two dimensional effective action reads
\be
I=\int d^2x\sqrt{-g}\left(\frac{1}{2}(\partial\varphi)^2+\frac{q}{4}\Phi_H\varphi {\cal R}+U(\varphi)\right).
\ee
On the other hand for the extremal black hole we are interested in the two dimensional part of the metric 
can be written as $ds^2=a^2(z)(-dt^2+dz^2)$ where $z=\int dr/a^2(r)$. It is easy to see that for the solutions
we have been discussing in which $a^2(r)$ has double zero at horizon, $z\rightarrow -\infty$ at $r\rightarrow
r_H$. Now the crucial observation \cite{{Solodukhin:1998tc},{Cvitan:2002cs}}
is that the trace of energy-momentum tensor of this two dimensional system  vanishes near the horizon and 
therefore the theory of the scalar field $\varphi$ approaches a CFT near the horizon with central charge 
$c=3\pi q^2\Phi_h^2$. The entropy of the black hole is also given by
\be
S=2\pi \Phi^2_H=\frac{\Omega_{D-2}}{4}\sum_{m=1}^{[D/2]}m\lambda_m\frac{(D-2)!}{(D-2m)!}\;
r_h^{D-2m}.
\ee
In particular for $D=5$ case one finds $S=\frac{\pi^2}{2}(r_H^3+12\lambda r_H)$ which upon using the
fact that in this case $r_H=V_{\rm eff}^{1/4}$ one get the same result as (\ref{5bh}). Writing the entropy in terms
of $\Phi_H$ is instructive. Actually $\Phi^2$ as a function of $r$ whose value at horizon gives the central
charge of the CFT, can be treated as a parameter which tells us how near we are to the horizon, where
the theory becomes a CFT. Therefore it is tempting to consider it as a C-function
\be
C=\Phi^2= \frac{\Omega_{D-2}}{8\pi}\sum_{m=1}^{[D/2]}m\lambda_m\frac{(D-2)!}{(D-2m)!}\;r^{D-2m}.
\ee
In fact for $D=4$ and in the leading order, $m=1$, it is the C-function introduced in \cite{Goldstein:2005rr}.
Using the same argument as that in \cite{Goldstein:2005rr} one can see that it is a
monotonically decreasing function. Moreover as we have seen at the fixed point (horizon) it is 
the central charge of the corresponding CFT theory. 

So far we have only considered the cases where the black hole has no electric charges. We note however that the
procedure can be generalized for the case where we have electric charges as well. For this case one may still
concentrate at the near horizon solution where  $r$ is kept fixed and indeed equal to the radius of the 
horizon and therefore the electric gauge fields can be chosen as $F^a_{rt}=\frac{4\pi}{\Omega_{D-2}}e^a$. Using the same
argument as \cite{Sen:2005wa} it is now easy to see that in this case the entropy can be found from the 
Legendre transformation of $f$ as follows
\be
S=\frac{4\pi}{a^2(r)''}(e^a\frac{\partial f}{\partial e^a}-f),
\label{LT}
\ee
and the electric charges of the black hole is given by $Q_a=\frac{\partial f}{\partial e^a}$ which for
our ansatz and the action (\ref{act}) it reads 
\be
Q_a=\frac{4\pi}{\Omega_{D-2}}\;\frac{v_2^{(D-2)/2}b^{D-2}}{v_1}f^{(e)}_{ab}e^b.
\ee
By making use of this expression and plugging it into the equation (\ref{LT}), one can see that
the entropy is given by equation (\ref{ENT}) with $f$ defined as (\ref{gb2}) but with a modified effective potential. 
The modified effective potential is given by 
\be
V_{\rm eff}=f^{(m)}_{\alpha\beta}p^\alpha p^\beta+\frac{2}{(D-2)!}{f^{(e)}}^{ab}Q_aQ_b\;.
\ee
where ${f^{(e)}}^{ab}$ is  the inverse of ${f^{(e)}}_{ab}$,
of course we assume that it is invertible. This procedure can also be used when we have axionic coupling as well.

Finally let us consider the five dimensional supersymmetric black hole studied in \cite{Strominger:1996sh} 
in more detail. The corresponding five dimensional action which can be obtained by compactification
of type II string theory on $K3\times S^1$ in the Einstein frame is given by
\be
I=\frac{1}{16\pi}\int d^4x\sqrt{-G}\left(R-\frac{4}{3}(\partial\phi)^2-\frac{e^{-4\phi/3}}{4}H^2
-\frac{e^{2\phi/3}}{4}F^2\right),
\ee
where $F$ is a RR 2-form field strength and $H$ is a 2-form axion field strength arising from the NS-NS 
3-form with one component tangent to the $S^1$. We shall consider the following ansatz for the near
horizon solution of the 5-dimensional black hole 
\bea
ds^2&=&v_1\left(-a^2(r)dt^2+\frac{dr^2}{a^2(r)}\right)+v_2r^2d\Omega_3,\;\;\;\;\;\;\;e^{2\phi/3}=u,\cr
F_{tr}&=&e_F,\;\;\;\;\;\;\;\;\;H_{tr}=\frac{4}{\pi}e_H,
\label{sol}
\eea
in which the entropy function, $f$, reads
\be
f=\frac{\pi}{8}v_1v_2^{3/2}r^3\left(\frac{6v_1-v_2r^2a^2(r)''}{v_1v_2r^2}+\frac{8e_H^2}{\pi^2v_1^2u^2}
+\frac{e^2_Fu}{2v_1^2}\right).
\ee
Therefore the black hole electric charges are given by
\be
Q_F=\frac{\partial f}{\partial e_F}=\frac{\pi}{8}\frac{v_2^{3/2}r^3}{v_1}ue_F,\;\;\;\;\;\;
Q_H=\frac{\partial f}{\partial e_H}=\frac{2}{\pi}\frac{v_2^{3/2}r^3}{v_1u^2}e_H,
\ee
Now one needs to extremize the entropy function with respect to $v_1,v_2$ and $u$. Doing so one finds
\be
v_1=\frac{a^2(r)''}{8}\left(\frac{8Q_HQ_F^2}{\pi^2}\right)^{1/3},\;\;\;\;\;
v_2=\frac{1}{r^2}\left(\frac{8Q_HQ_F^2}{\pi^2}\right)^{1/3},\;\;\;\;\;
u^3=\frac{1}{2}\left(\frac{4Q_F}{\pi Q_H}\right)^2,
\ee
which together with (\ref{LT}) can be used to find the entropy as follows  
\be
S=2\pi\sqrt{\frac{Q_HQ_F^2}{2}}.
\ee
The radius of horizon is also found as
\be
r_H=\left(\frac{8Q_HQ_F^2}{\pi^2}\right)^{1/6}.
\ee
 Essentially the result is the same as what we have considered in previous section up to a numerical factor. 
To understand the origin of this numerical 
mismatch one may use electric-magnetric duality to write 
the gauge fields as follows
\be
F_3=\frac{8}{\pi}Q_f\sqrt{\omega_3},\;\;\;\;\;\;\;\;\;H_3=2Q_H\sqrt{\omega_3}.
\ee
We note that in comparison with (\ref{solution}) one needs to rescale $p^\alpha$ proprely. Doing so
and going through the computations we find the correct numerical factors.  
Using the $R^2$ correction to the effective action one can also find the corrections to the entropy.

\noindent\textbf{Acknowledgments}

We would like to thank B. Chandrasekhar, A Ghodsi and S. Parvizi for discussions on the related topics.

\end{document}